\documentclass{article}

\usepackage[T1]{fontenc}
\usepackage[english]{babel}

\usepackage{lineno,hyperref}
\usepackage{color,graphicx,geometry,amsmath,commath,amssymb,bm,mathrsfs,textcomp}
\newcommand{\D}{\mathrm{d}}
\usepackage[font=footnotesize,skip=5pt]{caption}
\usepackage{enumitem,calc}
\usepackage[small]{titlesec}
\usepackage{hanging}

\usepackage{cite}
\usepackage[normalem]{ulem}

\title{Formation and morphology of closed and porous films grown from grains seeded on substrates: Two-dimensional simulations}
\author{Stoffel D.\ Janssens\footnote{Corresponding authors. \newline \hspace*{0.5cm} \textit{E-mail addresses:} stoffel.janssens@oist.jp (Stoffel D.\ Janssens), eliot.fried@oist.jp (Eliot Fried)}~, David V\'azquez-Cort\'es, and Eliot Fried$^{\text{\textasteriskcentered}}$}
\date{}

\begin{document}
\maketitle
\vspace{-1cm}
\begin{center}
Mechanics and Materials Unit (MMU), Okinawa Institute of Science and Technology Graduate University (OIST), 1919-1 Tancha, Onna-son, Kunigami-gun, Okinawa, Japan 904-0495
\end{center}
\vskip15pt
Two-dimensional simulations are used to explore topological transitions that occur during the formation of films grown from grains that are seeded on substrates. This is done for a relatively large range of the initial value $\Phi_s$ of the grain surface fraction $\Phi$. The morphology of porous films is captured at the transition when grains connect to form a one-component network using newly developed raster-free algorithms that combine computational geometry and network theory. Further insight on the morphology of porous films and their suspended counterparts is obtained by studying the pore surface fraction $\Phi_p$, the pore over grain ratio, the pore area distribution, and the contribution of pores of certain chosen areas to $\Phi_p$. Pinhole survival is evaluated at the transition when film closure occurs using survival function estimates. The morphology of closed films ($\Phi = 1$) is also characterized and is quantified by measuring grain areas and perimeters. The majority of investigated quantities are found to depend sensitively on $\Phi_s$ and the long-time persistence of pinholes exhibits critical behavior as a function of $\Phi_s$. In addition to providing guidelines for designing effective processes for manufacturing thin films and suspended porous films with tailored properties, this work may advance the understanding of continuum percolation theory.
\vskip5pt

\noindent
\emph{Keywords:} Grain growth, Chemical vapor deposition, Thin films, Microstructure, Voronoi diagram
\setcounter{footnote}{0}
\vskip10pt
\begin{center}
\line(1,0){450}
\end{center}
\vskip20pt

\section{Introduction}
The percolation threshold in continuum percolation theory \cite{Yi2004,Huang2021,Balberg2021} plays a significant role in thin film growth \cite{Amar1996}, grain boundary engineering \cite{Frary2005,Fullwood2006}, and research on porous media \cite{Sahimi1994,King2021}. In the formation of a thin film from a random distribution of initially isolated grains on a substrate, we identify the percolation threshold with the instant when the grains form a cluster that connects two opposing edges of the substrate. As a consequence of randomness, these instances differ for a comparable set of samples. We identify the percolation transition with the region spanned by the probability density of these instances. Inspired by network theory \cite{Barabasi2016}, we specify and explore two other topological thresholds that, to the best of our knowledge, have received little attention in the existing literature: the connected-grain threshold and the closed-film threshold. We identify the connected-grain threshold with the instant when all grains connect to form a one-component network and the closed-film threshold with the instant when the film closes. For a comparable set of samples, the probability densities of these respective instances span a connected-grain and a closed-film transition.

Granted that (1) grain nucleation occurs randomly and homogeneously over the entire substrate, (2) the growth rate is constant in time, (3) the growth is radial, and (4) grain motion \cite{Mullins1956,lazar2010,Saye2011} is negligible, the growth of thin films can be described by the Johnson--Mehl--Avrami--Kolmogorov (JMAK) model \cite{Johnson1939,Avrami1939,Shiryayev1992}. Despite its simplicities, the JMAK model can be applied to good effect in a plethora of circumstances \cite{Pineda1999,Jonas2009,Katsufuji2020}. An important outcome of the model is that the characteristic length of grains at film closure monotonically decreases as the ratio $n_r/g_r$ of the nucleation and growth rates $n_r$ and $g_r$ increases \cite{Moghadam2016}. This can be understood from realizing that the grain density $\rho$  increases monotonically as $n_r/g_r$ increases and that, at film closure, the characteristic length of grains and the film thickness decrease monotonically as $\rho$ increases. In general, the characteristic length after film closure is correlated with the thickness of the film depending on the dominant growth and restructuring mechanisms \cite{Janssens2011,Dulmaa2021}. A prime example of process for which $n_r/g_r$ is too small to form films in the submicron scale is provided by the chemical vapor deposition of polycrystalline diamond (PCD) \cite{Paritosh1999,Schreck2014,Stehlik2017,Janssens2020}. Nevertheless, films in the submicron scale can be grown by seeding nanometer scale diamond grains, known as nanodiamonds \cite{Ozawa2007,Mochalin2012,Sutisna2021}, on substrates prior to growth \cite{Williams2007,Tsigkourakos2012,Pobedinskas2021}. The presence of seeded grains before growth is not only beneficial for decreasing film thickness but can also facilitate the growth of films with desirable morphologies and properties \cite{Sarkar2018}. Moreover, altering the size of seeded nanodiamonds is known to strongly impact the morphology and thermal properties of gallium nitride (GaN) on PCD wafers \cite{Liu2017} and PCD on GaN wafers \cite{Smith2020}.

In this work, we investigate the formation and the morphology of closed and porous films grown from grains that are seeded on substrates using two-dimensional simulations. In so doing, our purpose is to provide guideline for designing effective processes for manufacturing thin films and suspended porous films with tailored properties. For example, suspended PCD films, which are typically made by removing a portion of a substrate upon a PCD film was grown, are prime candidates for single-cell culture and analysis \cite{Janssens2019}, particularly if they are porous. The morphological features of porous films are captured with vector-based algorithms that we developed combining computational geometry and network theory. Rasters-based methods \cite{Farjas2007}, which are prone to resolution issues, are not used. The specific goals of this work are to:
\begin{itemize}
\item Characterize the morphologic features of closed films, for which the grain surface fraction $\Phi$ is unity, by studying grain areas and perimeters.
\item Investigate the long-time persistence of pores at the closed-film transition. Such pores are typically referred to as pinholes.
\item Identify the connected-grain transition.
\item Characterize, at the connected-grain transition, the morphology of porous films and their suspended counter parts in terms of (1) the pore surface fraction $\Phi_p$, (2) the pore over grain ratio, (3) the pore area distribution, and (4) the contribution of pores of certain chosen areas to $\Phi_p$.
\item Ascertain the effect of the value $\Phi_s$ of the grain surface fraction at the outset of growth on the goals listed above.
\end{itemize}
The difference between porous films and their suspended counterparts is clarified hereinafter and the results of our study are summarized in the conclusions.

\section{Methods}
\label{Methods}
\subsection{Model}
The assumptions and simplifications underlying our approach resemble those of the JMAK model. The main difference is that nucleation of grains during growth is neglected. Instead, seeding is simulated with random sequential adsorption (RSA) \cite{Evans1993,Torquato2006}. During RSA, grain centers are sequentially placed at random positions on the substrate, with the provision that a grain is rejected if it overlaps one or more earlier adsorbed grains. For each simulation conducted, the number of deposited grains and the size of the square substrate are fixed, so that $\Phi$ is altered with the size of the deposited grains. RSA seeding is done with circular grains, namely disks, of fixed radius $r_s$ that expand radially and at an identical rate during growth. Consequently, grain boundaries correspond to a Voronoi diagram \cite{Okabe2000,Gavrilova2008,Aurenhammer2012}, which often describes the morphology of physical systems \cite{Finney1970,Sanchez-Gutierrez2016,Stemper2021}, and each grain fills a corresponding Voronoi cell at the instant of film closure.

\begin{figure}
\centering
\includegraphics{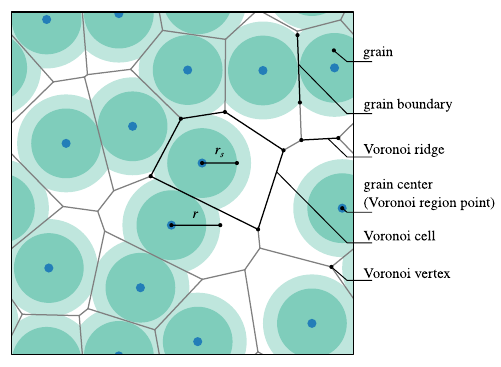}
\caption{Simulation model: Schematic depicting the ingredients of the model used in this work to simulate growth from grains that are seeded on substrates before growth. In this model, seeding is simulated by the random sequential adsorption (RSA) of circular grains, namely disks of fixed radius $r_s$, periodic boundary conditions are obeyed, and grains expand radially and at an identical rate. At the outset of growth, each grain is represented by a dark green disk and light green is used to indicate the instantaneous extent to which each grain has grown. The growth parameter $r$ defines the radius of a grain that is hypothetically unclustered. Grain centers act as Voronoi region points and define a Voronoi diagram in which ridges coincide with boundaries.}
\label{fig:model}
\end{figure}

Fig.~\ref{fig:model} provides a schematic depicting all ingredients of our model and the links between the morphology of a film and a Voronoi diagram. The figure also clarifies that grain boundaries are identified as grain-grain interfaces \cite{Stutton1995,Cantwell2014}. The growth parameter $r$ is defined as the radius of a grain that is hypothetically unclustered, and can be calculated through the relation
\begin{equation}
r(t) = r_s + g_r(t)t,
\label{Eq:r}
\end{equation}
in which $t$ is the growth time. From \eqref{Eq:r}, we find that the growth rate $g_r$ can vary with $t$. This fact can be crucial for modeling thermally activated growth processes, such as the chemical vapor deposition of diamond, that reach a constant temperature and, therefore, a constant growth rate, only after an initial stage in which transient effects are evident.

Unclustered diamond grains often take the shape of cuboctahedrons \cite{Bonnot1992}, for which spheres can serve as good approximations. The projection of a sphere onto the substrate is a disk. Therefore, disks can provide suitable representatives of grains in two-dimensional simulations. For a detailed geometrical modeling study on the chemical vapor deposition of single crystal diamond grains that includes the dependence of growth rate on facet type, we refer to the work of Silva \emph{et al.}\ \cite{Silva2008}.

\subsection{Seeding method}
In Uhlmann's \cite{Uhlmann2020} work on three-dimensional RSA of spheres, the standard deviation of the volume of a Voronoi cell approximately converges when simulations with $22^3$ spheres are performed. From this, we infer that for two-dimensional RSA of disks (grains), at least $22^2$ disks should be used for Voronoi cell properties to converge.

Seeding simulations are done by sequentially placing $10^4$ disks on a square substrate and for each simulation, $r_s$ and the area of the substrate are fixed. The initial value $\Phi_s$ of the grain surface fraction (or packing fraction) $\Phi$ is taken to range from $\Phi_s=0$ to $\Phi_s = 0.5184$ by altering $r_s$ and is limited by the saturation packing fraction $\Phi_s \approx 0.5471$ \cite{Zhang2013,Ciesla2018}. For each of the 19 $r_s$ values, $10^4$ simulations are performed. Periodic boundary conditions are implemented by surrounding the substrate with eight identical substrates on which disks are copied. These copied disks are not allowed to overlap the disks of neighboring substrates, which is probable when boundaries are crossed.

When $\Phi_s$ approached saturation, computation time significantly increased. This is due to the strongly decreasing disk adsorption near the end of the seeding simulation \cite{Pomeau1980}.

\subsection{Methods for obtaining thresholds}
\begin{figure}
\centering
\includegraphics{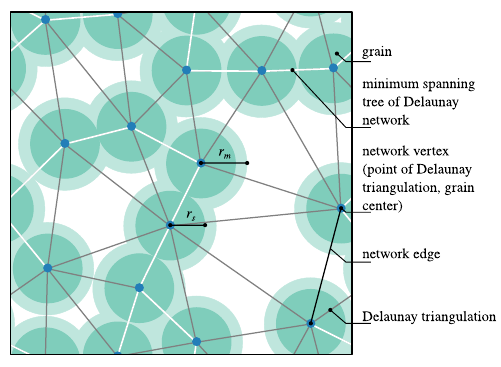}
\caption{Connected-grain threshold: A Delaunay network of 14 points that coincide with grain centers is depicted. The points and triangle edges correspond to network vertices and network edges, respectively, and the lengths of the triangle edges correspond to the network edge weights. Radius $r = r_m$, which marks the connected-grain threshold, is found with the minimum spanning (weight) tree (MST) of the Delaunay network by taking $2r_m$ to be the largest weight of the MST.}
\label{fig:mst}
\end{figure}

The radius $r = r_f$, which marks the closed-film threshold, is found by calculating the distances from Voronoi region points to neighboring Voronoi vertices and selecting the largest distance.

To find $r = r_m$, which marks the connected-grain threshold, a Delaunay triangulation of grain centers is transformed into an undirected network, as schematically depicted in Fig.~\ref{fig:mst}. Hereinafter, such a network is called a Delaunay network. The vertices and edges of a Delaunay network correspond to those of its Delaunay triangulation. The edges are appointed weights equal to the lengths of their corresponding triangle edges. The radius $r_m$ is then found via the minimum spanning (weight) tree (MST) of the underlying Delaunay network and is calculated using Kruskal's algorithm \cite{Kruskal1956}. We do this by taking $2r_m$ to be the largest weight of the MST.

For each value of $r_s$, the mean value $M_e[r_f]$ of $r_f$ and the mean value $M_e[r_m]$ of $r_m$ are obtained from the $10^4$ seeding simulations. We emphasize that the values of $M_e[r_f]$ and $M_e[r_m]$ increase as the number of grains in our simulations increases. However, investigating this dependence goes beyond the scope of our work. In this work, we investigate the trends obtained as a function of $r_s$. The probability densities (PDs) of $r_f$ and of $r_m$ span the closed-film and connected-grain transitions, respectively. 

Voronoi diagrams and Delaunay triangulations are made from grain centers, including the eight copies of these centers that are used for obeying the periodic boundary conditions. A more efficient procedure to construct Voronoi diagrams and Delaunay triangulations with periodic boundaries might be available in the near future \cite{Osang2020}.

\subsection{Methods for obtaining morphologies}
\label{Morphology}
\begin{figure*}
\centering
\includegraphics[width=\textwidth]{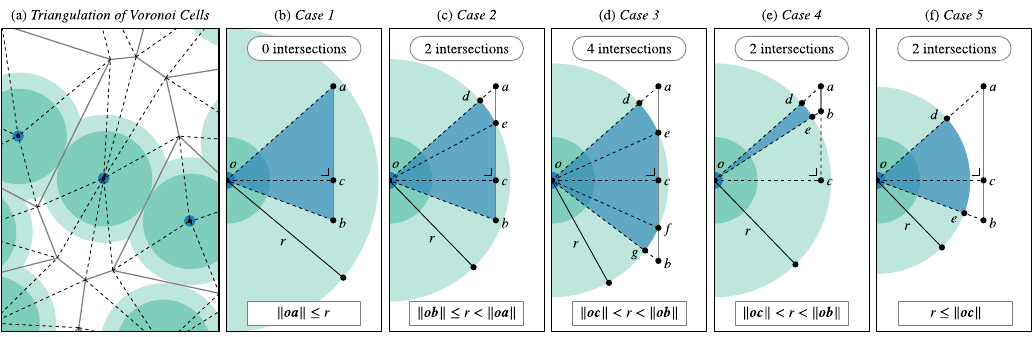}
\caption{Voronoi diagram triangulation and grain--triangle intersections: (a) A Voronoi diagram that is constructed from grain centers and for which the Voronoi cells are triangulated. For one Voronoi cell, the vertices of a triangle coincide with a grain center and the vertices of one Voronoi ridge. Grain surface fraction $\Phi$ is the sum of the areas of intersection of such triangles and grains. (b--f) Five cases where the perimeter of a grain, with center $o$ and radius $r$, and the edges of a triangle, with vertices $o$, $a$, and $b$, intersect. Intersections are denoted by the points $d$, $e$, $f$, and $g$. The number of intersections combined with the displayed relations specify each case. As delineated in the text, this forms the basis for the algorithms to calculate surface fractions and pore properties.}
\label{fig:intersection}
\end{figure*}

\subsubsection{Closed films}
\label{closed_films}
Grains of closed films correspond to Voronoi cells and when calculating the area and the perimeter of those grains, only grains with a region point lying on the original substrate are used. The periodic boundary conditions ensure that the total area of such cells is equal to the substrate area. Triangulation, as illustrated in Fig.~\ref{fig:intersection}(a), is used to calculate the area of a Voronoi cell. The perimeter of a Voronoi cell is calculated from its vertices.

\subsubsection{Algorithm 1: Pore surface fraction}
The pore surface fraction $\Phi_p$ is calculated by triangulating Voronoi cells in the same fashion as in Section~\ref{closed_films}. In each Voronoi cell, a grain center $o$ and the vertices $a$ and $b$ of a Voronoi ridge coincide with the vertices of a triangle. The grain surface fraction $\Phi$ is the sum of the areas of intersection of such triangles and grains, divided by the area of the substrate. Subsequently, $\Phi_p$ is calculated as $1 - \Phi$.

To compute the areas of intersection geometrically, we distinguish five cases for which the perimeter of a grain and a triangle intersect, as shown in Figs~\ref{fig:intersection}(b--f). The area of intersection $a_{in}$ for each of these cases is listed below.
\begin{itemize}
\item \emph{Case 1}: $a_{in}$ is equal to the area of triangle $oab$.
\item \emph{Case 2}: $a_{in}$ is equal to the sum of the area of triangle $oeb$ and the area of the sector $ode$ of the circle of radius $r$ centered at $o$.
\item \emph{Case 3}: $a_{in}$ is equal to the sum of the area of triangle area $oef$ and the areas of the sectors $ode$ and $ofg$ of the circle of radius $r$ centered at $o$.
\item \emph{Cases 4 \& 5}: $a_{in}$ is equal to the area of the sector $ode$ of the circle of radius $r$ centered at $o$.
\end{itemize}
Based on these cases, we developed a computational geometry algorithm that only uses angles when circle sectors are computed. The mathematics behind the most essential portion of the algorithm, which computes $a_{in}$, is given in the comments of our Python code. See S1 in supplementary materials for this code.

\subsubsection{Algorithm 2: Pore areas and pore perimeters}
We compute pore areas and pore perimeters using an algorithm that constructs a multi-component undirected network starting from a Voronoi network. The vertices and edges of this network correspond to those of a Voronoi diagram. The instructions of this algorithm, which are also generated by our Python code, are based on the five cases of intersection shown in Figs.~\ref{fig:intersection}(b--f) and are listed below.
\begin{itemize}
\item \emph{Case 1}: Voronoi network vertices $v_a$ and $v_b$, which correspond to Voronoi vertices $a$ and $b$, respectively, are filtered (removed) so that network edge $w$, formed between $v_a$ and $v_b$, vanishes.
\item \emph{Case 2}: Area $aed$ and perimeter portion $de$ are assigned to $v_a$. Vertex $v_b$ and edge $w$ are filtered.
\item \emph{Case 3}: Area $aed$ and perimeter portion $de$ are assigned to $v_a$. Area $bgf$ and perimeter portion $fg$ are assigned to $v_b$. Edge $w$ is filtered.
\item \emph{Cases 4 \& 5}: Area $abed$ and perimeter portion $de$ are arbitrarily assigned to $v_a$ or to $v_b$.
\end{itemize}
A non-filtered vertex is associated with multiple areas and the sum of these areas is the area weight of that vertex. An analogous requirement applies to perimeter portions. A component of the network, which consists of one or more vertices, represents a pore; the sum of the area weights of a pore equals the pore area $a_p$. The sum of the perimeter portion weights of that pore equals the perimeter $s_p$.

For a substrate with 14 grains, growth is shown in Figs.~\ref{fig:pores}(a--d) to showcase network formation. See S2 in supplementary materials for the corresponding movie. Light yellow, medium light orange, and medium dark pink  Voronoi vertices are contained in pores that cross boundaries in distinct ways and are treated to obey the periodic boundary conditions accordingly. Pores with dark purple colored vertices don't require such attention.
\begin{figure}
\centering
\includegraphics{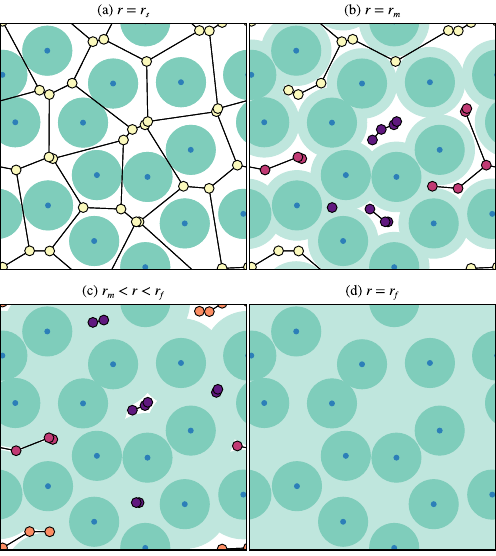}
\caption{Pore identification with \emph{Algorithm 2}: (a--d) A substrate seeded with 14 grains at several stages of growth with colored pore vertices. See S2 in supplementary materials for the corresponding movie. Pores with light yellow, medium light orange and medium dark pink colored vertices cross boundaries in distinct ways and are treated to obey periodic boundary conditions accordingly. The dark purple colored vertices don't require such attention. The algorithm of identifying pores and attributing pore properties to vertices, namely \emph{Algorithm 2}, is delineated in the text.}
\label{fig:pores}
\end{figure}
\begin{figure}
\centering
\includegraphics{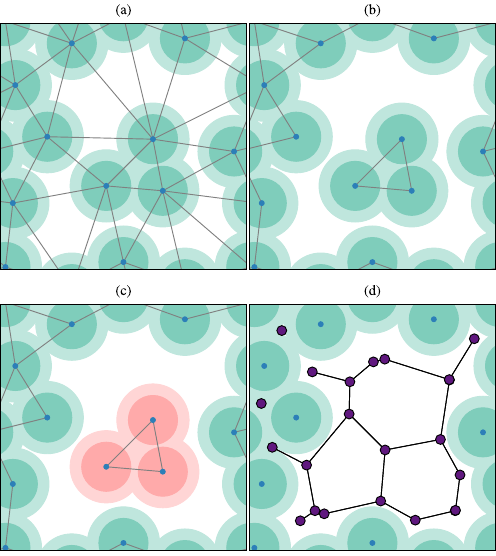}
\caption{Illustration of the film portion filtering (FPF) algorithm: (a) Delaunay network formed from grain centers with $r < r_m$. (b) Network with edges larger than $2r$ filtered. (c) All network components, except the component consisting of most vertices, are marked to be treated as special pore components. (d) Pores after FPF together with the vertices and edges obtained with \emph{Algorithm 2}.}
\label{fig:FPF}
\end{figure}

\subsubsection{Algorithm 3: Film portion filtering}
If the substrate of a porous film with $r < r_m$ is removed to make a suspended porous film, portions of film that are not connected to the sample spanning portion of film are removed together with the substrate. To simulate this process, we introduce a film portion filtering (FPF) algorithm that is illustrated in Figs. 5(a--c) and relies on constructing a Delaunay network of grain centers. Delaunay network edges that are larger than $2r$ are filtered and all network components, except the component consisting of most vertices, are marked to be treated as special pore components. These special components do not contribute to perimeter portion weights and have a maximal contribution to the area weights. Pores remaining after FPF are depicted in Fig.~\ref{fig:FPF}(d), together with the pore vertices and edges that are obtained via \emph{Algorithm 2}.

\subsection{Technical methods}
Simulations were done with Python. We also relied on the computational geometry software that we developed for computing grain-triangle surface intersections and for transforming Voronoi diagrams in pore networks, as described in Section \ref{Morphology}. This software was used in tandem with the graph-tool \cite{Peixoto2014} library, which allowed us to obtain MSTs and network components efficiently. In addition, the NumPy \cite{Harris2020}, SciPy, Matplotlib, and powerlaw \cite{Alstott2014} libraries were used. Voronoi diagrams were created using the implementation of Qhull \cite{Barber1996} in Scipy.

\subsection{Scaling}
\label{Scaling}
We adopt a scaling under which areas are multiplied by grain density $\rho$ so that lengths are multiplied by $\sqrt{\rho}$. This strategy can be rationalized by realizing that for a film formed by grains whose centers coincide with the vertices of a Bravais lattice, the area of a grain at film closure is exactly $1/\rho$. Following Tanemura \cite{Tanemura2003}, scaled perimeters are multiplied by $1/4$. The dimensionless quantities
\begin{equation}
A_c = a_c \rho
\qquad\text{and}\qquad
S_c = \frac{s_c\sqrt{\rho}}{4}
\label{eq:Aandb}
\end{equation}
represent the Voronoi cell area $a_c$ and perimeter $s_c$, respectively. As a consequence of these conventions, $S_c$ is unity for a Voronoi diagram that coincides with a square lattice. Following Pike and Seager \cite{Pike1974}, scaled radii are multiplied by $\sqrt{\pi}$, leading to dimensional quantity
\begin{equation}
R = r \sqrt{\pi \rho},
\label{scaled_r}
\end{equation}
which represents the scaled counterpart of $r$. Similarly, $R_s = r_s \sqrt{\pi \rho}$, $R_f = r_f \sqrt{\pi \rho}$, and $R_m = r_m \sqrt{\pi \rho}$ represent the scaled counterparts of $r_s$, $r_f$, and $r_m$. As a consequence of these conventions, $\Phi_s$ can be expressed as
\begin{equation}
\Phi_s = R_s^2.
\label{r_s_2_Phi_s}
\end{equation}
Another consequence of the chosen scaling is that the values of $R_s$ in this work range from 0 to 0.72.

\subsection{Towards experimental validation}
To experimentally validate the simulation results that follow, it would be feasible, for example, to examine future samples with one or more of the following techniques: (1) scanning electron microscopy (SEM), (2) electron backscatter diffraction (EBSD), (3) atomic force microscopy (AFM).
 
SEM and AFM can be used to obtain grain size distributions. AFM might give better results because it provides accurate height information. EBSD often requires polishing prior to measurements but can provide accurate grain size distributions.

SEM and AFM might render sufficient contrast between substrate and grains, making it possible to analyze pores and obtain pore size distributions.
\begin{figure*}
\centering
\includegraphics[width=\textwidth]{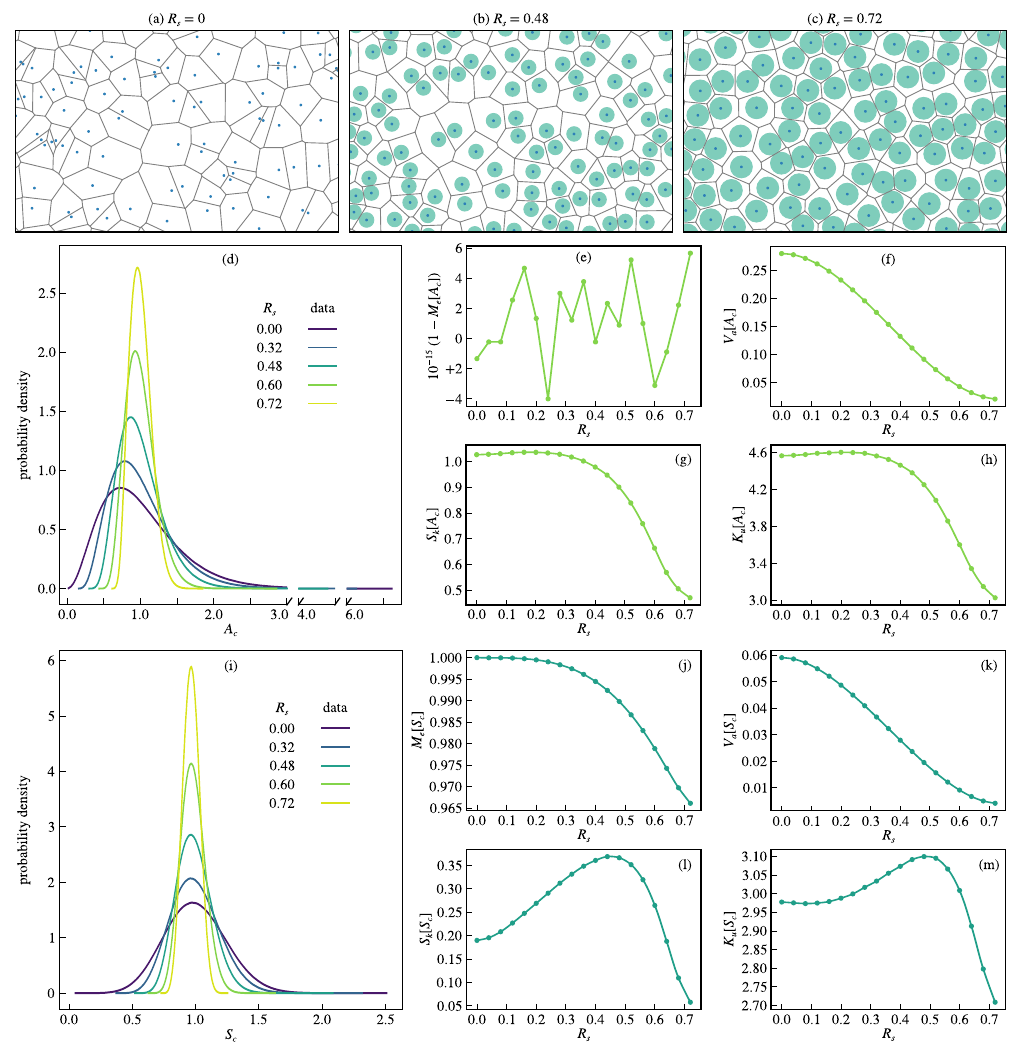}
\caption{Morphology of closed films: (a--c) Portions of a substrates seeded with grains of scaled radius $R_s = 0$, $R_s = 0.48$ and $R_s = 0.72$, respectively. Grain centers depicted as dots are used to construct Voronoi diagrams that represent the grain boundaries at film closure ($\Phi = 1$). (d--h) Probability densities (PDs) of $A_c$, $1 - M_e[A_c]$, variance $V_a[A_c]$, skewness $S_k[A_c]$, and kurtosis $K_u[A_c]$ for representative choices of $R_s$. Mean $M_e[A_c]$ is unity and $V_a[A_c]$ decreases as $R_s$ increases but $S_k[A_c]$ and $K_u[A_c]$ first increase. (i--m) The PDs of $S_c$, $M_e[S_c]$, $V_a[S_c]$, $S_k[S_c]$, and $K_u[S_c]$ for representative choices of $R_s$. $M_e[S_c]$ and $V_a[S_c]$ decrease monotonically as $R_s$ increases but $S_k[A_c]$ and $K_u[A_c]$ first increase. Results are discussed in the text.}
\label{fig:voronoi_results}
\end{figure*}
\section{Results and discussion}
\subsection{Morphology of closed films}
\label{Voronoi}
Portions of a substrate after seeding for representative choices of $R_s$ are shown in Figs.~\ref{fig:voronoi_results}(a--c), together with the Voronoi diagrams that are constructed from the grain centers. Each Voronoi cell represents a grain at film closure for which $\Phi = 1$. Since seeding is achieved by the Poisson point process \cite{Ferenc2007} for $R_s = 0$, the corresponding diagram is of Poisson--Voronoi type. With the aim of investigating the effect of $R_s$ on the morphology of the film by evaluating grain areas and perimeters through $A_c$ and $S_c$, we revisit and extend the results of Zhu \emph{et al.}\ \cite{Zhu2001}. In contrast to the scaling of Zhu \emph{et al.}\ \cite{Zhu2001}, the scaling detailed in Subsection~\ref{Scaling} allows us to explore how mean perimeter values vary with $R_s$. Inspired by the work of Torquato \emph{et al.}\ \cite{Torquato2021}, in which higher-order structural information is found to be of fundamental importance for characterizing density fluctuations, the work of Zhu \emph{et al.}\ \cite{Zhu2001} is further extended by investigating skewness $S_k[x]$ and kurtosis $K_u[x]$, which provide information on the asymmetry and tailedness of the PD of $x = \{A_c, S_c\}$, respectively. Here, $S_k[x]$ and $K_u[x]$ are defined as 
\begin{equation}
S_k[x]=\frac{\sum_{i=1}^{n}(x_i-M_e[x])^3}{nV_a[x]^{3/2}} 
\end{equation} 
and
\begin{equation}
K_u[x]=\frac{\sum_{i=1}^{n}(x_i-M_e[x])^4}{nV_a[x]^{2}},
\end{equation} 
respectively, where $x_i$, $i = 1,\dots, n$, denote the observed values of $x$ and $V_a[x]$ denotes the variance of $x$. 

Fig.~\ref{fig:voronoi_results}(d) shows the probability densities (PDs) of $A_c$ for representative values of $R_s$ and Figs.~\ref{fig:voronoi_results}(e--f) show how $1-M_e[A_c]$ and $V_a[A_c]$ vary with $R_s$. $M_e[A_c]$ is unity, as expected, and $V_a[A_c]$ decreases monotonically as $R_s$ increases, which is consistent with Figs.~\ref{fig:voronoi_results}(a--d). These results are in line with the literature \cite{Okabe2000,Zhu2001,Tanemura2003}. In Figs.~\ref{fig:voronoi_results}(g--h), $S_k[A_c]$ and $K_u[A_c]$ show a marginal increase from $R_s = 0$ to $R_s \approx 0.25$. For $R_s>0.3$, $S_k[A_c]$ and $K_u[A_c]$ decrease strongly.

Fig.~\ref{fig:voronoi_results}(i) shows the PDs of $S_c$ for the same choices of $R_s$ as in Fig.~\ref{fig:voronoi_results}(d) and Fig.~\ref{fig:voronoi_results}(j) depicts $M_e[S_c]$, which exhibits a monotonic reduction from unity with $R_s$. Plots provided in Figs.~\ref{fig:voronoi_results}(k--m) show that $V_a[S_c]$ decreases monotonically and that $S_k[S_c]$ and $K_u[S_c]$ first increase before decreasing.

The result for $M_e[S_c]$ can be understood from Fig.~\ref{fig:voronoi_results}(c), from which it is evident that configurations in which the grains exhibit nearly hexagonal order can occasionally occur, and from the work of Miles \cite{Miles1970}, who proved that $M_e[S_c] = 1$ for $R_s = 0$. The cell perimeter for the packing in Fig.~\ref{fig:voronoi_results}(c) is
\begin{equation}
S_h= \frac{\sqrt{3}}{2\sqrt{\cos(\pi/6)}}\approx0.931,
\label{hex_}
\end{equation}
which is smaller than that corresponding to square packing. This result also indicates that order increases with $R_s$, which can be expected since the rejection of grains by the RSA process, for $R_s > 0$, induces a correlation between grain center positions. Based on these findings, we are led naturally to propose the order metric
\begin{equation}
\mathcal{O} = \frac{1 - M_e[S_c]}{1 - S_h},
\label{eq:hex}
\end{equation}
which is zero for $R_s = 0$, approximately $0.49$ for $R_s = 0.72$, and unity for hexagonally packed grains. Although $\mathcal{O}$ does not detect order for square packings, it might provide useful measures of the degree of order in topological insulators constructed from random point sets \cite{Mitchell2018} and microfluidic pillar arrays that are placed on perturbed hexagonal lattices \cite{Walkama2020,Haward2021}. A critical review of various alternative measures of order is provided by Torquato \cite{Torquato2018}.

As $R_s$ increases, we hypothesize that the initial increase observed in $S_k[A_c]$ and $K_u[A_c]$ arises because the left tails of the PDs are restricted to $A_c\sim R_s^2$ and that the initial increase observed in $S_k[S_c]$ and $K_u[S_c]$ arises because the left tails of the PDs are restricted to $S_c\sim R_s$. Recognizing that the probability density functions (PDFs) of $A_c$ and $S_c$ for a hexagonal packing are Dirac delta functions, $V_a[A_c]$, $V_a[S_c]$, $S_k[A_c]$, $S_k[S_c]$, $K_u[A_c]$, and $K_u[S_c]$ should decrease as $R_s\to0.95$, which corresponds to a hexagonal grain packing. These expectations are confirmed by the plots in Figs.~\ref{fig:voronoi_results}(f--h) and Figs.~\ref{fig:voronoi_results}(k--m).

Inspired by the results obtained for $A_c$ and $S_c$, we hypothesize that $A_c$ and $S_c$ achieve particular maximum values at the saturation packing fraction. However, proving this assertion problem, which amounts to a problem in circle packing \cite{Stephenson2005}, goes beyond the scope of this work.
\subsection{Closed-film transition and pinhole survival}
\label{Pinholes}
To understand film closure, with the aim of guiding growth of films that are relatively thin but pinhole-free, we estimate the value of $R$ for which a low fraction of substrates contain pinholes. Here, we do this through the survival function estimate $P$ of $R_f$. $P = 10^{-3}$, for example, means that, on average, one out of $10^3$ substrates contain a film with at least one pinhole.

\begin{figure*}
\centering
\includegraphics[width=\textwidth]{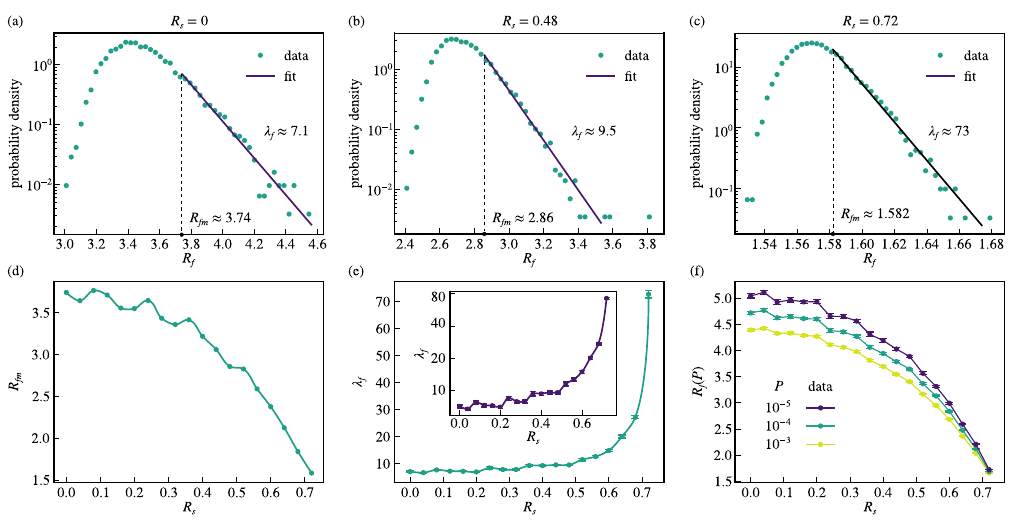}
\caption{Closed-film transition and pinhole survival: (a--c) Semi-logarithmic graphs of the probability densities (PDs) for representative choices of $R_f$. Based on a likelihood ratio test \cite{Clauset2009, Alstott2014} we approximate the PD tails with exponential distributions. All fits are obtained with \eqref{D} after estimating exponent $\lambda_f$ with \eqref{lambda}. $R_{fm}$, which denotes the value of $R_f$ from which exponential decay is assumed, is obtained from a Kolgomorov-Smirnov statistic \cite{Clauset2009}. (d) $R_{fm}$ displaying a monotonic decrease with $R_s$. (e) Values of $\lambda_f$, with error bars representing the standard deviation of $\lambda_f$. The inset is a semi-logarithmic graph of the same data, showing superexponential behavior. (f) $R = R_f$ for representative choices of $R_s$ and for $P = 10^{-3}$, $10^{-4}$, and $10^{-5}$, with $P$ denoting the chance that one film with $10^4$ grains contains pinholes. Results are discussed in the text.}
\label{fig:R_f}
\end{figure*}

The PDs of $R_f$ are depicted for representative choices of $R_s$ in Figs.~\ref{fig:R_f}(a--c). From the semi-logarithmic graphs in Figs.~\ref{fig:R_f}(a--c), we infer that these PDs exhibit the characteristics of exponential distributions for large values of $R_f$. However, from logarithmic plots not shown here, it might alternatively be expected that these PDs follow power-law distributions. Using a likelihood ratio test \cite{Clauset2009, Alstott2014}, we find that for the 19 simulations performed in this work, for which $R_s$ is systematically changed, 14 exhibit a preference to follow an exponential distribution. For $R_s =0.16$, $0.24$, $0.44$, $0.48$, and $0.56$, a power-law distribution is preferred. In what follows, we apply the most probable of the two types of decay, namely exponential decay. An important difference between both distributions is that exponential decay is asymptotically faster than power-law decay. This means that for exponential decay, pinhole free films can be thinner than for power-law decay.
On using $R_{fm}$ to denote the value of $R_f$ from which exponential decay is assumed, the exponent estimate $\lambda_f$ of an exponential distribution is calculated with the maximum likelihood fitting method as
\begin{equation}
\lambda_f = \frac{1}{M_e[R_f \geq R_{fm}] - R_{fm}},
\label{lambda}
\end{equation}
where $M_e[R_f \geq R_{fm}]$ denotes the mean of all $R_f$ values larger than or equal to $R_{fm}$. The quantity $R_{fm}$ is obtained with a Kolgomorov--Smirnov statistic \cite{Clauset2009} and $P$ is written as
\begin{equation}
P = Q e^{-\lambda_f(R_f - R_{fm})},
\label{P}
\end{equation}
with $Q$ being given by
\begin{equation}
Q = \frac{1}{n}\sum_{i = m}^{n} f(R_{fi}),
\label{Q}
\end{equation}
where $f(R_{fi}) = 1$ and $R_{fi}$, $i=1,\dots,m,\dots,n$, denote the observed values of $R_f$ sorted from small to large. We refer to Appendix~A for the derivation of \eqref{P}. By restructuring \eqref{P}, we obtain 
\begin{equation}
R_f = R_{fm} - \frac{1}{\lambda_f}\ln{\left(\frac{P}{Q}\right)},
\label{R_fP}
\end{equation}
for $R_f \geq R_{fm}$. The fits in Figs.~\ref{fig:R_f}(a--c) are obtained via a PDF estimate
\begin{equation}
D = Q \lambda_f e^{-\lambda_f(R_f - R_{fm})},
\label{D}
\end{equation}
which is also derived in Appendix~A. Fig.~\ref{fig:R_f}(d) shows that $R_{fm}$ decreases monotonically as $R_s$ increases and Fig.~\ref{fig:R_f}(e) shows that $\lambda_f$ increases monotonically as $R_s$ increases. Error bars represent the standard deviation of $\lambda_f$, which is approximated by $\lambda_f/\sqrt{n-m+1}$. For $R_s > 0.5$, critical behavior that is superexponential is clearly observable. We attribute this behavior to the fact that the $R_f$ PDF for $R_s\approx0.95$ (a hexagonal grain packing) is a Dirac delta function, for which $\lambda_f$ should be very large.

Fig.~\ref{fig:R_f}(f) shows that $R_f$ decreases monotonically, as $R_s$ increases, for $P=10^{-3}$, $10^{-4}$, and $10^{-5}$. For $P = 10^{-5}$, we assume that exponential decay persists well beyond the largest $R_f$ values obtained in our simulations. Fig.~\ref{fig:R_f}(f) also shows that the rate
\begin{equation}
\mathcal{V} = -\frac{\D R_f}{\D P},
\label{rate}
\end{equation}
depends on $R_s$. On combining \eqref{R_fP} and \eqref{rate}, this rate becomes
\begin{equation}
\mathcal{V} = \frac{1} {\lambda_fP}.
\label{V2}
\end{equation}

For a fixed grain density, the results in this section show that as $r_s$ increases, film closure occurs at lower values of $r$ and the long-time survival of pinholes critically diminishes.

\subsection{Connected-grain transition}
\label{porous_growth}
\begin{figure*}
\centering
\includegraphics[width=\textwidth]{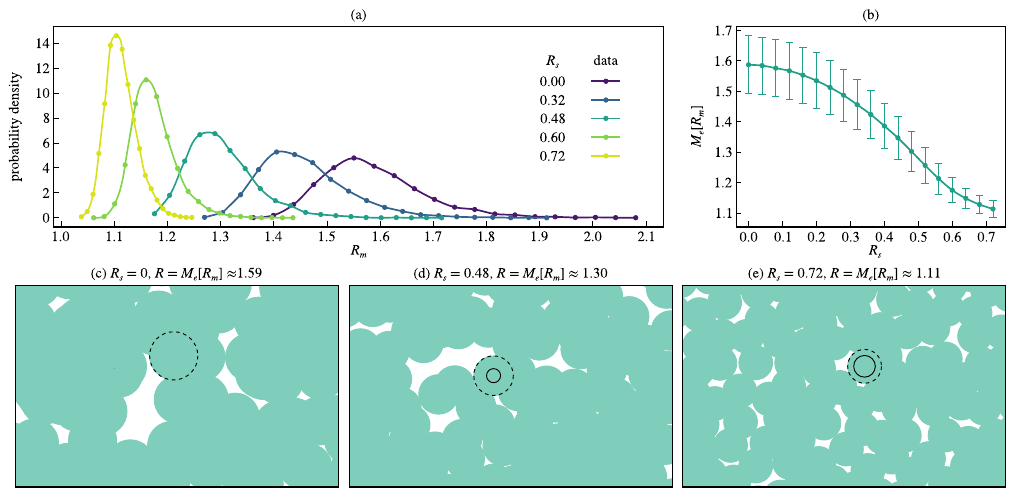}
\caption{Connected-grain transition: (a) $R_m$ PDs for representative choices of $R_s$ that are obtained from minimum spanning (weight) trees. (b) Mean values $M_e[R_m]$ versus $R_s$, with error bars representing the standard deviations of $R_m$. Both $M_e[R_m]$ and the standard deviation of $R_m$ decreases monotonically as $R_s$ increases. (c--e) The same portions of substrate as in Figs.~\ref{fig:voronoi_results}(a--c) with $R = M_e[R_m]$. The Voronoi diagrams, grain centers, and the locations of the grains before growth are suppressed in favor of pore visualization. Circles drawn with a dashed line have a radius $R = M_e[R_m]$ and circles drawn with a solid line have a radius of $R = R_s$. See S3, S4, and S5 in supplementary materials for figures similar to Figs.~\ref{fig:r_m}(c--e), respectively, but with $10^4$ grains. Results are discussed in the text.}
\label{fig:r_m}
\end{figure*}

The PDs of $R_m$ for representative choices of $R_s$ obtained from minimum spanning (weight) trees are depicted in Fig.~\ref{fig:r_m}(a). In Fig.~\ref{fig:r_m}(b), $M_e[R_m]$ is plotted versus $R_s$ together with error bars representing the standard deviations. Both $M_e[R_m]$ and the standard deviations decrease monotonically with $R_s$, as expected from Fig.~\ref{fig:r_m}(a). For a fixed grain density, these results show that, as $r_s$ increases, the formation of porous films occurs at lower values of $r$.

For a hexagonal lattice, $\Phi_s = \pi / 2\sqrt{3}$ and $R_m = \sqrt{\Phi_s} \approx 0.95$, which is below but relatively close to the value of $M_e[R_m]$ in Fig.~\ref{fig:r_m}(b) for $R_s = 0.72$. This indicates that, as $R_s$ increases, $M_e[R_m]$ tends to approach the value of a hexagonal lattice. The standard deviation also approaches the value of the hexagonal lattice, namely zero, as a function of $R_s$, and a similar observation applies to $R_f$. This is confirmed by the fact that for a hexagonal packing $R_f = 2R_s/\sqrt{3} \approx 1.10$, which is smaller but relatively close to the minimum value of $R_f$ for $R_s = 0.72$ found from Fig.~\ref{fig:R_f}(c).

The portions of the seeded substrates shown in Figs.~\ref{fig:voronoi_results}(a--c) are depicted again in Figs.~\ref{fig:r_m}(c--e), with the difference that $R = M_e[R_m]$ and that Voronoi diagrams, grains centers, and the locations of grains before growth are suppressed in favor of pore visualization. Circles formed by dashed lines have radius $R = M_e[R_m]$ and circles formed by solid lines have radius $R = R_s$. See S3, S4, and S5 in supplementary materials for figures similar to Figs.~\ref{fig:r_m}(c--e), respectively, but with $10^4$ grains. From Figs.~\ref{fig:r_m}(c--e), we infer that the pore over grain ratio $H_p$ increases with $R_s$ and that the mean pore area $M_e[A_p]$ decreases with $R_s$. However, from the images it is unclear how $R_s$ affects the pore surface fraction $\Phi_p$.

\subsection{Morphology of porous films}
\begin{figure*}
\centering
\includegraphics[width=\textwidth]{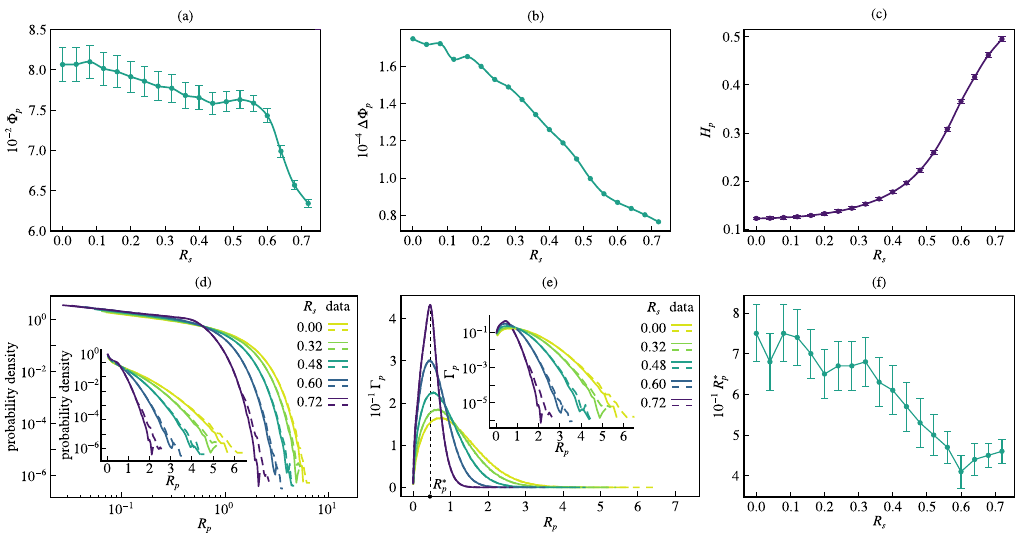}
\caption{Morphology of porous films: (a) Pore surface fraction $\Phi_p$ as a function of $R_s$. (b) Graph of $\Phi_p$ with film portion filtering (FPF) minus $\Phi_p$ without FPF versus $R_s$, showing a small dependence of $\Phi_p$ on FPF. (c) Pore over grain ratio $H_p$ shows a strong dependence on $R_s$. (d) Probability density of equivalent pore radius $R_p$ for representative choices of $R_s$ plotted as logarithmic and semi-logarithmic graphs. Dashed lines are plots including FPF. (e) Contribution $\Gamma_p$, obtained with \eqref{eq:pore_cont_dis}, of pores with a certain value of $R_p$ to $\Phi_p$, plotted as logarithmic and semi-logarithmic graphs for representative choices of $R_s$. Dashed lines are plots including FPF. $R_p = R_p^*$ denotes where the maximum value of $\Gamma_p$ is found for $R_s = 0.72$ and FPF. (f) Plot of $R_p^*$ versus $R_s$, for FPF, showing a decrease from approximately 0.75 to 0.45. Results are discussed in the text.}
\label{fig:pore_results}
\end{figure*}
In Fig.~\ref{fig:pore_results}(a), $\Phi_p$ is plotted as a function of $R_s$ for the case of film portion filtering (FPF), as described in Figs.~\ref{fig:pores}(e--g), with error bars representing the standard deviation of $\Phi_p$. $\Phi_p$ decreases from approximately $0.080$ to $0.064$, with an increase in intensity when $R_s > 0.55$. The values of $\Phi_p$ and corresponding values of $R_s$ for a square ($\Phi_p=1-\pi/4\approx 0.21$, $R_s=0.89$) and hexagonal ($\Phi_p=1-\pi/2\sqrt{3})\approx 0.093$, $R_s=0.95$) packings exhibit trends similar to those in Fig.~\ref{fig:pore_results}(a). This observation might not be coincidental since the results of Section \ref{Voronoi} demonstrate that grain related properties approach those expected for square and hexagonal packings depending on whether $R_s$ is large or small. From Fig.~\ref{fig:pore_results}(b), it is evident that the difference $\Delta\Phi_p$ between $\Phi_{pa}$ (obtained after FPF) and $\Phi_{pb}$ (obtained before FPF) is relatively small and decreases monotonically  as $R_s$ increases. This quantity can be regarded as the fraction of film that is lost during the process of making a suspended film. It should tend to zero at $R_s\approx0.95$, where the grains are in a hexagonal packing.

Fig.~\ref{fig:pore_results}(c) shows that the pore over grain ratio $H_p$ increases monotonically with $R_s$, which confirms the qualitative observations made from Figs.~\ref{fig:r_m}(c--e) in the last paragraph of Section \ref{porous_growth}. The trend might be governed by the structural ordering induced by the RSA process, since $H_p = 1$ for grains laying on perfectly ordered lattices. Figs.~\ref{fig:pores}(e--h) clarify that FPF has no effect on $H_p$.

On defining the equivalent pore radius $ R_p = \sqrt{A_p}$, Fig.~\ref{fig:pore_results}(d) shows a logarithmic graph of the PD of $R_p$ for representative values of $R_s$. Dashed lines correspond to values obtained for FPF and the inset shows a semi-logarithmic graph of the same PDs. As expected, larger values of $R_p$ are found for FPF. Furthermore, all PDs monotonically decay with $R_p$. From the inset, we infer that tails of the PDs show a decay that is at very least exponential and increases in intensity with $R_s$.

The contribution of pores with a certain area $A_p$ to $\Phi_p$, or similarly, the contribution $\Gamma_p$ of pores with a certain equivalent radius $R_p$ to $\sqrt{\Phi_p}$ is now investigated. We express $\Gamma_p$ via the relation
\begin{equation}
\Gamma_p = \frac{ \sqrt{\Phi_p} R_p G}{\sum{R_p G \Delta R_p}},
\label{eq:pore_cont_dis}
\end{equation}
where $\Delta R_p$ and $G$ denote the width of a data bin and the probability density of $R_p$, respectively, with $G$ as plotted in Fig.~\ref{fig:pore_results}(d). The sum is taken over all data bins and multiplication by $\sqrt{\Phi_p}$ ensures that differences in $\sqrt{\Phi_p}$ are incorporated when comparing porous films. In Fig.~\ref{fig:pore_results}(e), $\Gamma_p$ is plotted for representative choices of $R_s$. Dashed lines correspond to values obtained for FPF and the inset shows a semi-logarithmic graph of the same data. In contrast to $G$, $\Gamma_p$ drops to zero for $R_s=0$. Assuming that $\Gamma_p(R_s)$ is a continuous function close to $R_s=0$, this is consistent with physical expectations, since pores with a radius equal to zero should not contribute to $\Gamma_p$. The width of the peaks become more narrow with increasing $R_s$. This trend can be explained on recognizing that for grains in a hexagonal packing, $\Gamma_p$ is represented by a Dirac delta function. The values of $R_p=R_p^*$, which represent the pore areas that contribute most to $\Phi_p$, are found at the maxima of $\Gamma_p$ and are plotted versus $R_s$ in Fig.~\ref{fig:pore_results}(f). Each value of $R_p^*$ is obtained by binning a data set in $100\pm1$ bins and selecting the bin containing the largest number of data points. An error bar represents the width of a bin. As a function of $R_s$, $R_p^*$ decreases from approximately $0.75$ to $0.45$. These observations are indistinguishable from those obtained without FPF, which shows that FPF has a relatively small impact on $R_p^*$.

\section{Conclusions}
We investigate the formation and the morphology of closed and porous films that are grown from grains seeded on substrates through two-dimensional simulations. The purpose is to provide guideline for designing effective processes for manufacturing thin films and suspended porous films with tailored properties. Simulations are done for a relatively large range of the initial value $\Phi_s$ of the grain surface fraction $\Phi$. The assumptions and simplifications underlying our model resemble those of the Johnson--Mehl--Avrami--Kolmogorov model. The main difference is that nucleation of grains during growth is neglected. Instead, grains are seeded with circular grains of radius $r_s$ that are not allowed to overlap. For all simulations, the number of deposited grains and the size of the square substrate are fixed, so that $\Phi_s$ is altered with $r_s$. A growth parameter $r$ is defined as the radius of a grain that is hypothetically unclustered. The morphology of porous films is found by introducing raster-free algorithms that are based on computational geometry and networks theory. Our findings are summarized below:

\begin{itemize}

\item For closed films ($\Phi=1$), we evaluate the effect of the initial grain surface fraction $\Phi_s$ on the morphology through grain areas and perimeters. In contrast to earlier work, we adopt a scaling which affords comparisons of the mean value of the grain perimeters. As $r_s$ increases, that quantity is found to decrease towards the value that is expected for the hexagonal packing of grains. Based on this result, we determine a simple metric for order that can be useful for describing amorphous or perturbed hexagonal systems. Furthermore, the skewness and the kurtosis of the grain area and grain perimeter probability densities are found to first increase and then decrease as $r_s$ increases. The initial increase is reasoned to be caused by the lower bound that $r_s$ sets on the size of a grain.

\item We evaluate film closure with the aim of guiding the growth of films that are relatively thin but pinhole-free. We find that the mean value $M_e[r_f]$ of $r=r_f$, at which film closure occurs, decreases monotonically as $r_s$ increases. For all values of $r_s$, film closure is found to be linked to a process with features resembling those of exponential decay. We also find that the long-time persistence of pinholes critically decreases as $r_s$ increases.

\item The mean value $M_e[r_m]$ of $r = r_m$, at which all grains connect to form a one-component network, is found to decrease monotonically as $r_s$ increases. For $r = M_e[r_m]$, we also find that the pore surface fraction $\Phi_p$ monotonically decreases, the pore over grain ratio strongly increases, and the pore area that contributes most to $\Phi_p$ decreases as $r_s$ increases. The latter quantities are also obtained for porous films that are made suspended by substrate removal, with the aim of guiding the fabrication of such films. If $r < r_m$, small portions of film that are not connected to the sample spanning film are removed together with the substrate, which enlarges the size of pores. We find that the evaluated quantities remain practically identical after substrate removal.

\item Although the results in this work can, to the best of our knowledge, not be obtained analytically, we find that trends obtained as $r_s$ increases can to some degree be explained by asymptotic analysis. This is done by calculating the exact values of several quantities for hexagonally packed grains for which $\Phi_s \approx 0.91$.

\end{itemize}
A scaling is used such that results can be interpreted as a function of $r_s$ or as a function of grain density, however, we underline that using another number of grains per simulation leads to different results. We leave the investigation of this dependency to future work.

\section*{Data availability}
Further data and codes of this study are available per reasonable request.

\section*{Declaration of competing interest}
The authors declare that they have no known competing financial interests or personal relationships that could have appeared to influence the work reported in this paper.

\section*{Acknowledgements}
We gratefully acknowledge support from the Okinawa Institute of Science and Technology Graduate University with subsidy funding from the Cabinet Office, Government of Japan, and thank Martin Skrodzki for suggesting the use of minimum spanning trees. S.\ D.\ J.\ and E.\ F.\ also acknowledge Kakenhi funding from the Japan Society for the Promotion of Science [grant number 21K03782].

\appendix
\section*{Appendix A. Survival function estimate}
\setcounter{equation}{0}
\renewcommand{\theequation}{A.\arabic{equation}}

If the empirical (cumulative) distribution function $F$ of $R_f$, written as 
\begin{equation}
F = \frac{1}{n}\sum_{i = 1}^{m} f(R_{fi}),
\label{F_a}
\end{equation}
in which $f(R_{fi}) = 1$ and  $R_{fi}$, $i = 1,\dots,m,\dots,n$, denotes the observed values of $R_f$ sorted from small to large, then it follows that
\begin{equation}
\frac{1}{n}\sum_{i = 1}^{m - 1} f(R_{fi}) + \frac{1}{n}\sum_{i = m}^{n} f(R_{fi}) 
=\frac{1}{n}\sum_{i = 1}^{m - 1} f(R_{fi}) +Q
= 1,
\label{F_1_a}
\end{equation}
where $Q$ is the partial sum defined in \eqref{Q}. If $R_{fm}$ denotes the value from which the probability density function corresponding to $F$ is estimated by $D$, then $Q$ can be approximated by
\begin{equation}
Q\approx \int_{R_{fm}}^{\infty} D\,\D R_f.
\label{Q_a_1}
\end{equation}
On taking $D$ to be of the exponential form
\begin{equation}
D = C e^{-\lambda_f R_f},
\label{D_a}
\end{equation}
where $C$ denotes a scale factor. From \eqref{Q_a_1} and \eqref{D_a}, $C$ is given by
\begin{equation}
C = Q \lambda_f e^{\lambda_f R_{fm}}.
\label{C_a}
\end{equation}
Substituting \eqref{C_a} in \eqref{D_a} yields
\begin{equation}
D = Q \lambda_f e^{-\lambda_f(R_f - R_{fm})}.
\label{D_e_a}
\end{equation}
On using \eqref{D_e_a} in \eqref{Q_a_1}, it follows from \eqref{D_a} and \eqref{F_1_a} that
\begin{eqnarray}
F & \approx & \frac{1}{n}\sum_{i = 1}^{m - 1} f(R_{fi}) + Q \int_{R_{fm}}^{R_f} \lambda_f e^{-\lambda(R_f - R_{fm})} \D R_f\\ 
& \approx & 1 - Q e^{-\lambda_f(R_f - R_{fm})}.
\end{eqnarray}
Finally, if a survival function is defined as the difference between unity and the salient empirical distribution function, then the relation for survival function estimate $P$ is
\begin{equation}
P = Q e^{-\lambda_f(R_f - R_{fm})}.
\label{P_a}
\end{equation}

\section*{Appendix B. Supplementary materials}


\begin{thebibliography}{10}
\expandafter\ifx\csname url\endcsname\relax
  \def\url#1{\texttt{#1}}\fi
\expandafter\ifx\csname urlprefix\endcsname\relax\def\urlprefix{URL }\fi
\expandafter\ifx\csname href\endcsname\relax
  \def\href#1#2{#2} \def\path#1{#1}\fi

\bibitem{Yi2004}
Y.-B. Yi, A.~M. Sastry, Analytical approximation of the percolation threshold
  for overlapping ellipsoids of revolution, Proc. R. Soc. London, Ser. A 460
  (2004) 2353--2380.
\newblock \href {http://dx.doi.org/10.1098/rspa.2004.1279}
  {\path{doi:10.1098/rspa.2004.1279}}.

\bibitem{Huang2021}
X.~Huang, D.~Yang, Z.~Kang, Impact of pore distribution characteristics on
  percolation threshold based on site percolation theory, Phys. A 570 (2021)
  125800.
\newblock \href {http://dx.doi.org/10.1016/j.physa.2021.125800}
  {\path{doi:10.1016/j.physa.2021.125800}}.

\bibitem{Balberg2021}
I.~Balberg, Principles of the theory of continuum percolation, in: M.~Sahimi,
  A.~G. Hunt (Eds.), Complex media and percolation theory, 1st Edition,
  Encyclopedia of complexity and systems science series, Springer, New York,
  NY, 2021, pp. 89--148.
\newblock \href {http://dx.doi.org/10.1007/978-1-0716-1457-0_95}
  {\path{doi:10.1007/978-1-0716-1457-0_95}}.

\bibitem{Amar1996}
J.~G. Amar, F.~Family, Kinetics of submonolayer and multilayer epitaxial
  growth, Thin Solid Films 272 (1996) 208--222.
\newblock \href {http://dx.doi.org/10.1016/0040-6090(95)06947-X}
  {\path{doi:10.1016/0040-6090(95)06947-X}}.

\bibitem{Frary2005}
M.~Frary, C.~A. Schuh, Grain boundary networks: scaling laws, preferred cluster
  structure, and their implications for grain boundary engineering, Acta Mater.
  53 (2005) 4323--4335.
\newblock \href {http://dx.doi.org/10.1016/j.actamat.2005.05.030}
  {\path{doi:10.1016/j.actamat.2005.05.030}}.

\bibitem{Fullwood2006}
D.~T. Fullwood, J.~A. Basinger, B.~L. Adams, Lattice-based structures for
  studying percolation in two-dimensional grain networks, Acta Mater. 54 (2006)
  1381--1388.
\newblock \href {http://dx.doi.org/10.1016/j.actamat.2005.11.012}
  {\path{doi:10.1016/j.actamat.2005.11.012}}.

\bibitem{Sahimi1994}
M.~Sahimi, Applications of percolation theory, 1st Edition, Taylor \& Francis,
  1994.
\newblock \href {http://dx.doi.org/10.1201/9781482272444}
  {\path{doi:10.1201/9781482272444}}.

\bibitem{King2021}
P.~King, M.~Masihi, Percolation in porous media, in: M.~Sahimi, A.~G. Hunt
  (Eds.), Complex media and percolation theory, 1st Edition, Encyclopedia of
  complexity and systems science series, Springer, New York, NY, 2021, pp.
  237--254.
\newblock \href {http://dx.doi.org/10.1007/978-1-0716-1457-0_389}
  {\path{doi:10.1007/978-1-0716-1457-0_389}}.

\bibitem{Barabasi2016}
A.~Barab\'asi, M.~P\'osfai, Network science, 1st Edition, Oxford Press Press,
  2016.

\bibitem{Mullins1956}
W.~W. Mullins, Two-dimensional motion of idealized grain boundaries, J. Appl.
  Phys. 27 (1956) 900--904.
\newblock \href {http://dx.doi.org/10.1063/1.1722511}
  {\path{doi:10.1063/1.1722511}}.

\bibitem{lazar2010}
E.~A. Lazar, R.~D. MacPherson, D.~J. Srolovitz, A more accurate two-dimensional
  grain growth algorithm, Acta Mater. 58 (2010) 364--372.
\newblock \href {http://dx.doi.org/10.1016/j.actamat.2009.09.008}
  {\path{doi:10.1016/j.actamat.2009.09.008}}.

\bibitem{Saye2011}
R.~I. Saye, J.~A. Sethian, The {V}oronoi implicit interface method for
  computing multiphase physics, Proc. Natl. Acad. Sci. U. S. A. 108 (2011)
  19498--19503.
\newblock \href {http://dx.doi.org/10.1073/pnas.1111557108}
  {\path{doi:10.1073/pnas.1111557108}}.

\bibitem{Johnson1939}
W.~A. Johnson, R.~F. Mehl, Reaction kinetics in processes of nucleation and
  growth, Trans. Am. Inst. Min. Metall. Pet. Eng. 135 (1939) 410--458.

\bibitem{Avrami1939}
M.~Avrami, Kinetics of phase change. {I} {G}eneral theory, J. Chem. Phys. 7
  (1939) 1103--1112.
\newblock \href {http://dx.doi.org/10.1063/1.1750380}
  {\path{doi:10.1063/1.1750380}}.

\bibitem{Shiryayev1992}
A.~N. Shiryayev, On the statistical theory of metal crystallization, in: A.~N.
  Shiryayev (Ed.), Selected works of {A}. {N}. {K}olmogorov, 1st Edition,
  Vol.~26 of Mathematics and its applications ({S}oviet series), Springer,
  {D}ordrecht, 1992, pp. 188--192.
\newblock \href {http://dx.doi.org/10.1007/978-94-011-2260-3_22}
  {\path{doi:10.1007/978-94-011-2260-3_22}}.

\bibitem{Pineda1999}
E.~Pineda, D.~Crespo, Microstructure development in {K}olmogorov,
  {J}ohnson-{M}ehl, and {A}vrami nucleation and growth kinetics, Phys. Rev. B
  60 (1999) 3104--3112.
\newblock \href {http://dx.doi.org/10.1103/PhysRevB.60.3104}
  {\path{doi:10.1103/PhysRevB.60.3104}}.

\bibitem{Jonas2009}
J.~J. Jonas, X.~Quelennec, L.~Jiang, E.~Martin, The {A}vrami kinetics of
  dynamic recrystallization, Acta Mater. 57 (2009) 2748--2756.
\newblock \href {http://dx.doi.org/10.1016/j.actamat.2009.02.033}
  {\path{doi:10.1016/j.actamat.2009.02.033}}.

\bibitem{Katsufuji2020}
T.~Katsufuji, T.~Kajita, S.~Yano, Y.~Katayama, K.~Ueno, Nucleation and growth
  of orbital ordering, Nat. Commun. 11 (2020) 2324.
\newblock \href {http://dx.doi.org/10.1038/s41467-020-16004-2}
  {\path{doi:10.1038/s41467-020-16004-2}}.

\bibitem{Moghadam2016}
M.~M. Moghadam, P.~W. Voorhees, Thin film phase transformation kinetics: from
  theory to experiment, Scr. Mater. 124 (2016) 164--168.
\newblock \href {http://dx.doi.org/10.1016/j.scriptamat.2016.07.010}
  {\path{doi:10.1016/j.scriptamat.2016.07.010}}.

\bibitem{Janssens2011}
S.~D. Janssens, P.~Pobedinskas, J.~Vacik, V.~Petr{\'{a}}kov{\'{a}}, B.~Ruttens,
  J.~{D'Haen}, M.~Nesl{\'{a}}dek, K.~Haenen, P.~Wagner, Separation of intra-
  and intergranular magnetotransport properties in nanocrystalline diamond
  films on the metallic side of the metal--insulator transition, New J. Phys.
  13 (2011) 083008.
\newblock \href {http://dx.doi.org/10.1088/1367-2630/13/8/083008}
  {\path{doi:10.1088/1367-2630/13/8/083008}}.

\bibitem{Dulmaa2021}
A.~Dulmaa, F.~G. Cougnon, R.~Dedoncker, D.~Depla, On the grain size-thickness
  correlation for thin films, Acta Mater. 212 (2021) 116896.
\newblock \href {http://dx.doi.org/10.1016/j.actamat.2021.116896}
  {\path{doi:10.1016/j.actamat.2021.116896}}.

\bibitem{Paritosh1999}
Paritosh, D.~J. Srolovitz, C.~Battaile, X.~Li, J.~Butler, Simulation of faceted
  film growth in two-dimensions: Microstructure, morphology and texture, Acta
  Mater. 47 (1999) 2269--2281.
\newblock \href {http://dx.doi.org/10.1016/S1359-6454(99)00086-5}
  {\path{doi:10.1016/S1359-6454(99)00086-5}}.

\bibitem{Schreck2014}
M.~Schreck, J.~Asmussen, S.~Shikata, J.-C. Arnault, N.~Fujimori, Large-area
  high-quality single crystal diamond, MRS Bull. 39 (2014) 504–510.
\newblock \href {http://dx.doi.org/10.1557/mrs.2014.96}
  {\path{doi:10.1557/mrs.2014.96}}.

\bibitem{Stehlik2017}
S.~Stehlik, M.~Varga, P.~Stenclova, L.~Ondic, M.~Ledinsky, J.~Pangrac,
  O.~Vanek, J.~Lipov, A.~Kromka, B.~Rezek, Ultrathin nanocrystalline diamond
  films with silicon vacancy color centers via seeding by 2 nm detonation
  nanodiamonds, ACS Appl. Mater. Interfaces 9 (2017) 38842--38853.
\newblock \href {http://dx.doi.org/10.1021/acsami.7b14436}
  {\path{doi:10.1021/acsami.7b14436}}.

\bibitem{Janssens2020}
S.~D. Janssens, B.~Sutisna, A.~Giussani, J.~A. Kwiecinski,
  D.~V\'azquez-Cort\'es, E.~Fried, Boundary curvature effect on the wrinkling
  of thin suspended films, Appl. Phys. Lett. 116 (2020) 193702.
\newblock \href {http://dx.doi.org/10.1063/5.0006164}
  {\path{doi:10.1063/5.0006164}}.

\bibitem{Ozawa2007}
M.~Ozawa, M.~Inaguma, M.~Takahashi, F.~Kataoka, A.~Kr\"{u}ger, E.~\={O}sawa,
  Preparation and behavior of brownish, clear nanodiamond colloids, Adv. Mater.
  19 (2007) 1201--1206.
\newblock \href {http://dx.doi.org/10.1002/adma.200601452}
  {\path{doi:10.1002/adma.200601452}}.

\bibitem{Mochalin2012}
V.~N. Mochalin, O.~Shenderova, D.~Ho, Y.~Gogotsi, The properties and
  applications of nanodiamonds, Nat. Nanotechnol. 7 (2012) 11--23.
\newblock \href {http://dx.doi.org/10.1038/nnano.2011.209}
  {\path{doi:10.1038/nnano.2011.209}}.

\bibitem{Sutisna2021}
B.~Sutisna, S.~D. Janssens, A.~Giussani, D.~V\'azquez-Cort\'es, E.~Fried, Block
  copolymer–nanodiamond coassembly in solution: Towards multifunctional
  hybrid materials, Nanoscale 13 (2021) 1639--1651.
\newblock \href {http://dx.doi.org/10.1039/D0NR07441A}
  {\path{doi:10.1039/D0NR07441A}}.

\bibitem{Williams2007}
O.~A. Williams, O.~Douh\'{e}ret, M.~Daenen, K.~Haenen, E.~\={O}sawa,
  M.~Takahashi, Enhanced diamond nucleation on monodispersed nanocrystalline
  diamond, Chem. Phys. Lett. 445 (2007) 255--258.
\newblock \href {http://dx.doi.org/10.1016/j.cplett.2007.07.091}
  {\path{doi:10.1016/j.cplett.2007.07.091}}.

\bibitem{Tsigkourakos2012}
M.~Tsigkourakos, T.~Hantschel, S.~D. Janssens, K.~Haenen, W.~Vandervorst,
  Spin-seeding approach for diamond growth on large area silicon-wafer
  substrates, Phys. Status Solidi A 209 (2012) 1659--1663.
\newblock \href {http://dx.doi.org/10.1002/pssa.201200137}
  {\path{doi:10.1002/pssa.201200137}}.

\bibitem{Pobedinskas2021}
P.~Pobedinskas, G.~Degutis, W.~Dexters, J.~{D'Haen}, M.~K. {Van Bael},
  K.~Haenen, Nanodiamond seeding on plasma-treated tantalum thin films and the
  role of surface contamination, Appl. Surf. Sci. 538 (2021) 148016.
\newblock \href {http://dx.doi.org/10.1016/j.apsusc.2020.148016}
  {\path{doi:10.1016/j.apsusc.2020.148016}}.

\bibitem{Sarkar2018}
R.~Sarkar, J.~Rajagopalan, Synthesis of thin films with highly tailored
  microstructures, Mater. Res. Lett. 6 (2018) 398--405.
\newblock \href {http://dx.doi.org/10.1080/21663831.2018.1471420}
  {\path{doi:10.1080/21663831.2018.1471420}}.

\bibitem{Liu2017}
D.~Liu, D.~Francis, F.~Faili, C.~Middleton, J.~Anaya, J.~W. Pomeroy, D.~J.
  Twitchen, M.~Kuball, Impact of diamond seeding on the microstructural
  properties and thermal stability of {GaN}-on-diamond wafers for high-power
  electronic devices, Scr. Mater. 128 (2017) 57--60.
\newblock \href {http://dx.doi.org/10.1016/j.scriptamat.2016.10.006}
  {\path{doi:10.1016/j.scriptamat.2016.10.006}}.

\bibitem{Smith2020}
E.~J.~W. Smith, A.~H. Piracha, D.~Field, J.~W. Pomeroy, G.~R. Mackenzie,
  Z.~Abdallah, F.~C.-P. Massabuau, A.~M. Hinz, D.~J. Wallis, R.~A. Oliver,
  M.~Kuball, P.~May, Mixed-size diamond seeding for low-thermal-barrier growth
  of {CVD} diamond onto {GaN} and {AlN}, Carbon 167 (2020) 620--626.
\newblock \href {http://dx.doi.org/10.1016/j.carbon.2020.05.050}
  {\path{doi:10.1016/j.carbon.2020.05.050}}.

\bibitem{Janssens2019}
S.~D. Janssens, D.~V\'azquez-Cort\'es, A.~Giussani, J.~A. Kwiecinski, E.~Fried,
  Nanocrystalline diamond-glass platform for the development of
  three-dimensional micro- and nanodevices, Diamond Relat. Mater. 98 (2019)
  107511.
\newblock \href {http://dx.doi.org/10.1016/j.diamond.2019.107511}
  {\path{doi:10.1016/j.diamond.2019.107511}}.

\bibitem{Farjas2007}
J.~Farjas, P.~Roura, Numerical model of solid phase transformations governed by
  nucleation and growth: {M}icrostructure development during isothermal
  crystallization, Phys. Rev. B 75 (2007) 184112.
\newblock \href {http://dx.doi.org/10.1103/PhysRevB.75.184112}
  {\path{doi:10.1103/PhysRevB.75.184112}}.

\bibitem{Evans1993}
J.~W. Evans, Random and cooperative sequential adsorption, Rev. Mod. Phys. 65
  (1993) 1281--1329.
\newblock \href {http://dx.doi.org/10.1103/RevModPhys.65.1281}
  {\path{doi:10.1103/RevModPhys.65.1281}}.

\bibitem{Torquato2006}
S.~Torquato, O.~U. Uche, F.~H. Stillinger, Random sequential addition of hard
  spheres in high euclidean dimensions, Phys. Rev. E 74 (2006) 061308.
\newblock \href {http://dx.doi.org/10.1103/PhysRevE.74.061308}
  {\path{doi:10.1103/PhysRevE.74.061308}}.

\bibitem{Okabe2000}
A.~Okabe, B.~Boots, K.~Sugihara, S.~N. Chiu, Spatial tessellations: {C}oncepts
  and applications of {V}oronoi diagrams, 2nd Edition, Wiley, 2000.

\bibitem{Gavrilova2008}
M.~L. Gavrilova, Generalized {V}oronoi diagram: A geometry-based approach to
  computational intelligence, 1st Edition, Springer-Verlag Berlin Heidelberg,
  2008.
\newblock \href {http://dx.doi.org/10.1007/978-3-540-85126-4}
  {\path{doi:10.1007/978-3-540-85126-4}}.

\bibitem{Aurenhammer2012}
F.~Aurenhammer, R.~Klein, D.~Lee, Voronoi Diagrams and Delaunay Triangulations,
  World Scientific, 2012.
\newblock \href {http://dx.doi.org/10.1142/8685} {\path{doi:10.1142/8685}}.

\bibitem{Finney1970}
J.~L. Finney, J.~D. Bernal, Random packings and the structure of simple
  liquids. {I}. {T}he geometry of random close packing, Proc. R. Soc. London,
  Ser. A 319 (1970) 479--493.
\newblock \href {http://dx.doi.org/10.1098/rspa.1970.0189}
  {\path{doi:10.1098/rspa.1970.0189}}.

\bibitem{Sanchez-Gutierrez2016}
D.~S{\'a}nchez-Guti{\'e}rrez, M.~Tozluoglu, J.~D. Barry, A.~Pascual, Y.~Mao,
  L.~M. Escudero, Fundamental physical cellular constraints drive
  self-organization of tissues, EMBO J. 35 (2016) 77--88.
\newblock \href {http://dx.doi.org/10.15252/embj.201592374}
  {\path{doi:10.15252/embj.201592374}}.

\bibitem{Stemper2021}
L.~Stemper, M.~A. Tunes, P.~Dumitraschkewitz, F.~Mendez-Martin, R.~Tosone,
  D.~Marchand, W.~A. Curtin, P.~J. Uggowitzer, S.~Pogatscher, Giant hardening
  response in {A}l{M}g{Z}n({C}u) alloys, Acta Mater. 206 (2021) 116617.
\newblock \href {http://dx.doi.org/10.1016/j.actamat.2020.116617}
  {\path{doi:10.1016/j.actamat.2020.116617}}.

\bibitem{Stutton1995}
A.~P. Sutton, R.~W. Balluffi, Interfaces in crystalline materials, 1st Edition,
  Oxford Press Press, 1995.

\bibitem{Cantwell2014}
P.~R. Cantwell, M.~Tang, S.~J. Dillon, J.~Luo, G.~S. Rohrer, M.~P. Harmer,
  Grain boundary complexions, Acta Mater. 62 (2014) 1--48.
\newblock \href {http://dx.doi.org/10.1016/j.actamat.2013.07.037}
  {\path{doi:10.1016/j.actamat.2013.07.037}}.

\bibitem{Bonnot1992}
A.~Bonnot, B.~Mathis, J.~Mercier, J.~Leroy, J.~Vitton, Growth mechanisms of
  diamond crystals and films prepared by chemical vapor deposition, Diamond
  Relat. Mater. 1 (1992) 230--234.
\newblock \href {http://dx.doi.org/10.1016/0925-9635(92)90030-R}
  {\path{doi:10.1016/0925-9635(92)90030-R}}.

\bibitem{Silva2008}
F.~Silva, X.~Bonnin, J.~Achard, O.~Brinza, A.~Michau, A.~Gicquel, Geometric
  modeling of homoepitaxial {CVD} diamond growth: {I}. {T}he
  \{100\}\{111\}\{110\}\{113\} system, J. Cryst. Growth 310 (2008) 187--203.
\newblock \href {http://dx.doi.org/10.1016/j.jcrysgro.2007.09.044}
  {\path{doi:10.1016/j.jcrysgro.2007.09.044}}.

\bibitem{Uhlmann2020}
M.~Uhlmann, Voronoi tessellation analysis of sets of randomly placed
  finite-size spheres, Phys. A 555 (2020) 124618.
\newblock \href {http://dx.doi.org/10.1016/j.physa.2020.124618}
  {\path{doi:10.1016/j.physa.2020.124618}}.

\bibitem{Zhang2013}
G.~Zhang, S.~Torquato, Precise algorithm to generate random sequential addition
  of hard hyperspheres at saturation, Phys. Rev. E 88 (2013) 053312.
\newblock \href {http://dx.doi.org/10.1103/PhysRevE.88.053312}
  {\path{doi:10.1103/PhysRevE.88.053312}}.

\bibitem{Ciesla2018}
M.~Cie\'sla, R.~M. Ziff, Boundary conditions in random sequential adsorption,
  J. Stat. Mech.{:} Theory Exp. 2018 (2018) 043302.
\newblock \href {http://dx.doi.org/10.1088/1742-5468/aab685}
  {\path{doi:10.1088/1742-5468/aab685}}.

\bibitem{Pomeau1980}
Y.~Pomeau, Some asymptotic estimates in the random parking problem, J. Phys. A:
  Math. Gen. 13 (1980) L193--L196.
\newblock \href {http://dx.doi.org/10.1088/0305-4470/13/6/006}
  {\path{doi:10.1088/0305-4470/13/6/006}}.

\bibitem{Kruskal1956}
J.~B. Kruskal, On the shortest spanning subtree of a graph and the traveling
  salesman problem, Proc. Amer. Math. Soc. 7 (1956) 48--50.
\newblock \href {http://dx.doi.org/10.1090/S0002-9939-1956-0078686-7}
  {\path{doi:10.1090/S0002-9939-1956-0078686-7}}.

\bibitem{Osang2020}
G.~Osang, M.~Rouxel-Labb{\'e}, M.~Teillaud, Generalizing {CGAL} periodic
  {D}elaunay triangulations, in: 28th Annual European Symposium on Algorithms
  (ESA 2020), Vol. 173 of Leibniz International Proceedings in Informatics
  (LIPIcs), Schloss Dagstuhl--Leibniz-Zentrum f{\"u}r Informatik, 2020, pp.
  75:1--75:17.
\newblock \href {http://dx.doi.org/10.4230/LIPIcs.ESA.2020.75}
  {\path{doi:10.4230/LIPIcs.ESA.2020.75}}.

\bibitem{Peixoto2014}
T.~P. Peixoto, The graph-tool python library, figshare. \href
  {http://dx.doi.org/10.6084/m9.figshare.1164194}
  {\path{doi:10.6084/m9.figshare.1164194}}.

\bibitem{Harris2020}
C.~R. Harris, K.~J. Millman, S.~J. van~der Walt, R.~Gommers, P.~Virtanen,
  D.~Cournapeau, E.~Wieser, J.~Taylor, S.~Berg, N.~J. Smith, R.~Kern, M.~Picus,
  S.~Hoyer, M.~H. van Kerkwijk, M.~Brett, A.~Haldane, J.~Fern{\'a}ndez~del
  R{\'i}o, M.~Wiebe, P.~Peterson, P.~G{\'e}rard-Marchant, K.~Sheppard,
  T.~Reddy, W.~Weckesser, H.~Abbasi, C.~Gohlke, T.~E. Oliphant, Array
  programming with {N}um{P}y, Nature 585 (2020) 357--362.
\newblock \href {http://dx.doi.org/10.1038/s41586-020-2649-2}
  {\path{doi:10.1038/s41586-020-2649-2}}.

\bibitem{Alstott2014}
J.~Alstott, E.~Bullmore, D.~Plenz, powerlaw: {A} {P}ython package for analysis
  of heavy-tailed distributions, {PL}o{S} {ONE} 9 (2014) e85777.
\newblock \href {http://dx.doi.org/10.1371/journal.pone.0085777}
  {\path{doi:10.1371/journal.pone.0085777}}.

\bibitem{Barber1996}
C.~B. Barber, D.~P. Dobkin, H.~Huhdanpaa, The quickhull algorithm for convex
  hulls, ACM Trans. Math. Softw. 22 (1996) 469--483.
\newblock \href {http://dx.doi.org/10.1145/235815.235821}
  {\path{doi:10.1145/235815.235821}}.

\bibitem{Tanemura2003}
M.~Tanemura, Statistical distributions of {P}oisson {V}oronoi cells in two and
  three dimensions, Forma 18 (2003) 221--247.

\bibitem{Pike1974}
G.~E. Pike, C.~H. Seager, Percolation and conductivity: {A} computer study.
  {I}, Phys. Rev. B 10 (1974) 1421--1434.
\newblock \href {http://dx.doi.org/10.1103/PhysRevB.10.1421}
  {\path{doi:10.1103/PhysRevB.10.1421}}.

\bibitem{Ferenc2007}
J.~Ferenc, Z.~N\'eda, On the size distribution of {P}oisson {V}oronoi cells,
  Phys. A 385 (2007) 518--526.
\newblock \href {http://dx.doi.org/10.1016/j.physa.2007.07.063}
  {\path{doi:10.1016/j.physa.2007.07.063}}.

\bibitem{Zhu2001}
H.~X. Zhu, S.~M. Thorpe, A.~H. Windle, The geometrical properties of irregular
  two-dimensional {V}oronoi tessellations, Philos. Mag. A 81 (2001) 2765--2783.
\newblock \href {http://dx.doi.org/10.1080/01418610010032364}
  {\path{doi:10.1080/01418610010032364}}.

\bibitem{Torquato2021}
S.~Torquato, J.~Kim, M.~A. Klatt, Local number fluctuations in hyperuniform and
  nonhyperuniform systems: {H}igher-order moments and distribution functions,
  Phys. Rev. X 11 (2021) 021028.
\newblock \href {http://dx.doi.org/10.1103/PhysRevX.11.021028}
  {\path{doi:10.1103/PhysRevX.11.021028}}.

\bibitem{Miles1970}
R.~E. Miles, On the homogeneous planar {P}oisson point process, Math. Biosci. 6
  (1970) 85--127.
\newblock \href {http://dx.doi.org/10.1016/0025-5564(70)90061-1}
  {\path{doi:10.1016/0025-5564(70)90061-1}}.

\bibitem{Mitchell2018}
N.~P. Mitchell, L.~M. Nash, D.~Hexner, A.~M. Turner, W.~T.~M. Irvine, Amorphous
  topological insulators constructed from random point sets, Nat. Phys. 14
  (2018) 380--385.
\newblock \href {http://dx.doi.org/10.1038/s41567-017-0024-5}
  {\path{doi:10.1038/s41567-017-0024-5}}.

\bibitem{Walkama2020}
D.~M. Walkama, N.~Waisbord, J.~S. Guasto, Disorder suppresses chaos in
  viscoelastic flows, Phys. Rev. Lett. 124 (2020) 164501.
\newblock \href {http://dx.doi.org/10.1103/PhysRevLett.124.164501}
  {\path{doi:10.1103/PhysRevLett.124.164501}}.

\bibitem{Haward2021}
S.~J. Haward, C.~C. Hopkins, A.~Q. Shen, Stagnation points control chaotic
  fluctuations in viscoelastic porous media flow, Proc. Natl. Acad. Sci. U. S.
  A. 118.
\newblock \href {http://dx.doi.org/10.1073/pnas.2111651118}
  {\path{doi:10.1073/pnas.2111651118}}.

\bibitem{Torquato2018}
S.~Torquato, Perspective: {B}asic understanding of condensed phases of matter
  via packing models, J. Chem. Phys. 149 (2018) 020901.
\newblock \href {http://dx.doi.org/10.1063/1.5036657}
  {\path{doi:10.1063/1.5036657}}.

\bibitem{Stephenson2005}
K.~Stephenson, Introduction to circle packing. {T}he theory of discrete
  analytic functions, 1st Edition, Cambridge University Press, 2005.

\bibitem{Clauset2009}
A.~Clauset, C.~R. Shalizi, M.~E.~J. Newman, Power-law distributions in
  empirical data, SIAM Rev. 51 (2009) 661--703.
\newblock \href {http://dx.doi.org/10.1137/070710111}
  {\path{doi:10.1137/070710111}}.

\end{thebibliography}
\end{document}